\documentstyle[twoside,fleqn,espcrc2,epsfig]{article}

\newcommand{\AmS}{{\protect\the\textfont2
  A\kern-.1667em\lower.5ex\hbox{M}\kern-.125emS}}

\title{Pure SU(3) Potentials}

\author{Sedigheh Deldar \thanks{Poster presented at the 17th International
Symposium, {\it Lattice99}, Pisa, June 29th to July 3rd, 1999.}
\address{Department of Physics, 
        Washington University, St. Louis, MO 63130, USA \\ }}
       
\begin{document}

\begin{abstract}
The string tensions for fundamental and a variety of representations:
adjoint, 6, 10, 15-symmetric, 15-antisymmetric, and 27
have been measured in pure gauge SU(3). The calculations have been done
on anisotropic lattices, using an $O(a^2)$ tadpole improved action.
A range of lattice spacings and lattice sizes has been used to control 
finite volume and finite lattice spacing effects.
Potentials between quarks in various representations have also been
calculated with the fat center vortices model.
At intermediate distances, the results
show approximate Casimir scaling. From my lattice calculations, no color 
screening has been observed so far, even for distances as large as about 
2.5 fm.
 
\end{abstract}

\maketitle


One of the main goals of a non-perturbative formulation of gauge theories
is to understand the phenomenon of confinement. The formation of a flux
tube and the linear confinement of static quarks in the fundamental
representation has been well established by pure gauge lattice QCD.
The confinement of static sources in higher representations of QCD is still
a question. Some numerical calculations have been done for SU(2) 
\cite{Bern82} and SU(3) \cite{Fabe88} at non-zero temperature. At 
zero temperature, there have been studies for 8 (adjoint) \cite{Camp86},
and 6 and 8 \cite{Mich98}. At {\it Lattice98}, I gave some 
preliminary results of string tensions of some higher representations
at zero temperature \cite{Deld99}. Independent of this work, at
{\it Lattice99}, G. Bali reported some results of static source
potentials of some higher representations \cite{Bali99}.

In this paper, I give results of measuring the potentials
and string tensions of static sources of fundamental and some higher
representations (6, 8, $15_{a}$, 10, 27, $15_{s}$) in pure SU(3).
Each representation can be labeled by the ordered pair $(n,m)$, with
n and m the original number of 3 and $\bar{3}$ which participated in
constructing the representation.
Screening is expected to occur for representations with zero triality:
$8 \equiv (1,1)$, $10 \equiv (3,0)$, and $27 \equiv (2,2)$. (Triality 
is defined as (n-m) mod 3.) For these representations, as the distance 
between the two adjoint
sources increases, the potential energy of the flux tube rises. We expect 
a pair of gluons to pop out of vacuum when this energy is equal or greater
than twice the glue-lump mass. For large distances, 
the static sources combine with the octet(8) charges (dynamic gluons) 
popped out of the vacuum and produce singlets which screen.
Therefore the potential between static sources is no longer
$R$ dependent. Static sources in representations with non-zero triality, 
$6 \equiv (2,0)$, $15_{a} \equiv (2,1)$ and $15_{s} \equiv (4,0) $, 
transform into the lowest order representation (3 and $\bar{3}$) 
by binding to the gluonic $8's$ which are popped out of the vacuum.
As a result, the slope of the linear potentials of the representations
with non-zero triality changes to the slope of the fundamental one,
and a universal string tension is expected for large distances.

In this work, the potential, $V(r)$, has been found by measuring Wilson loops. At large 
$t$, $W(r,t)\simeq \exp[-V(r)t]$ where $W(r,t)$ is the Wilson loop as a 
function of $r$, the spatial separation of the quark, and the 
propagation time $t$. 
Wilson loops of higher representations have been found in terms of Wilson
loops in the fundamental representation \cite{Deld99}. To calculate the string
tensions, the potentials obtained from Wilson loops have been fitted
to a linear plus Coulombic form.

\begin{figure}[htb]
\vspace{35pt}
\epsfxsize=1. \hsize
\epsffile{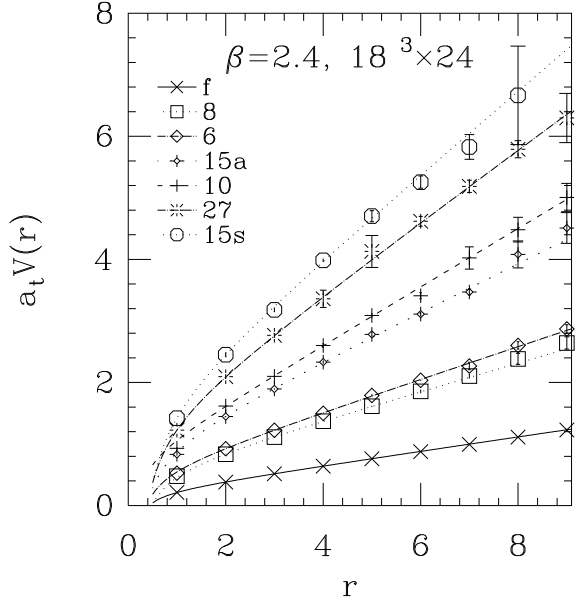}
\vspace{-61pt}
\caption{A typical plot of V(r) versus r for various representation of 
SU(3). $a_{s}\approx .25$ fm.}
\vspace{-25pt}
\label{fig:toosmall}
\end{figure}

\begin{figure}[htb]
\vspace{35pt}
\epsfxsize=1. \hsize
\epsffile{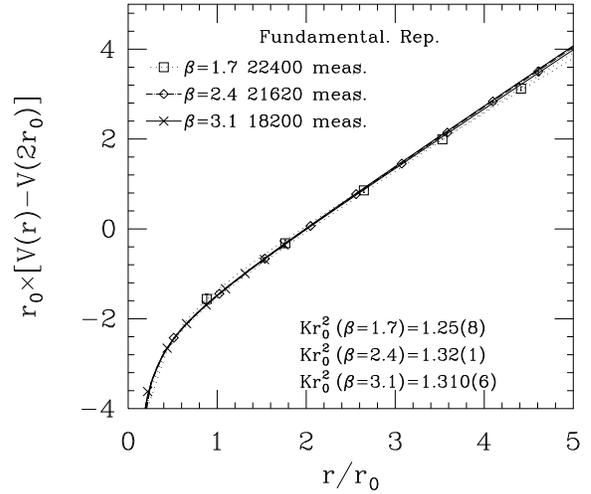}
\vspace{-60pt}
\caption{The static quark potential $V(R)$ in terms of hadronic scale
$r_{0}$ for fundamental representation. Points are the results of three
different measurements with coupling constants equal to 1.7, 2.4 and 3.1.}
\vspace{-20pt}
\label{fig:largenenough}
\end{figure}

The tadpole-improved tree level action of ref.\cite{Morn97} is used.
This action has $a_{t}\neq a_{s}$, where $a_{t}$, $a_{s}$ are the temporal
and spatial lattice spacings, respectively.
For $a_{t} \ll a_{s}$ the discretization error is of order
$O(a_{s}^4,a_{t}^2,a_{t}a_{s}^2)$.
For Wilson loops, smearing followed by projection back to SU(3) is 
performed on spatial links.

Measurements are done on four lattices: $10^3\times24$,
$8^3\times24$, $18^3\times24$, and $16^3\times24$ at $\beta$ equal to
$1.7,~ 2.4,~ 2.4$ and $3.1$, with aspect ratios of $5,3,3,$ and $1.5$,
respectively. The temporal lattice spacings are kept the same.
Coupling constants have been chosen based on the
approximate desired lattice spacing ratios: 4:2:1.
Fig 1 shows a typical plot of $V(r)$ versus $r$ for various 
representations on the $18^3\times24$ lattice at $\beta=2.4$. The
string tensions are qualitatively in agreement with Casimir scaling.
No color screening or change of the slope is observed for the adjoint or
other representations. (Screening should start at about
1.2 fm for the adjoint representation. \cite{Mich98})

To study the scaling behavior, the potential between the static sources
is found in terms of hadronic scale, $r_{0}$, 
($[r^2dV/dr]_{r=r_{0}}=1.65$), where $V$ is the potential between quarks 
in the fundamental representation. In Fig 2, the potential between two 
sources in fundamental representation in terms of $r_{0}$ is plotted for 
different lattice measurements. Good scaling behavior is observed for the 
fundamental representation. The scaling gets worse for higher 
representations, roughly in proportion to the string tension. 
Table 1 shows the best estimate for the string tension of each 
representation obtained by the weighted average of the four 
lattice measurements. The errors on the string tensions 
are the statistical error (from the weighted average), and the systematic error 
of discretization (determined by the standard deviation of the results over
the 3 couplings), and the error on $r_{0}$, respectively. 

I also examined
the fat-center-vortices model \cite{Debb97} to study its prediction
for the potentials. In this model,
the vacuum is a condensate of vortices 
of finite thickness. Confinement is produced by the 
independent fluctuations of the vortices piercing each unit area of a
Wilson loop.
The average Wilson loop in SU(N),
predicted by this model has the form:
\vspace{-5pt}
\begin{equation}
<W(C)> = \prod \{ 1 - \sum^{N-1}_{n=1} f_{n} (1 - Re {\cal G}_{r}
          [\vec{\alpha}^n_{C}(x)])\},
\end{equation}
\vspace{-5pt}
where
${\cal G}_{r}[\vec{\alpha}] = \frac{1}{d_{r}} Tr \exp[i\vec{\alpha} . \vec{H}].$
$d_{r}$ is the dimension of representation r, and
$\{H_{i}\}$ is the subset of the generators needed to generate
the center of the group.
$\alpha_{C}(x)$ depends on the
vortex location, and $f_{n}$ represents the probability that any 
given unit area is ``pierced" by a vortex.  
I have used the same function for $\alpha$ as Greensite {\it et al.}
used but with the appropriate normalization factor for SU(3). I
have found $\{H_{i}\}$ for each representation in SU(3).
Fig 3 represents the potentials between the static sources
obtained by this model. For
$R>40$, potentials for representations with zero triality are screened
and the ones with non-zero triality closely parallel the fundamental
potential. There exists a region, $R<20$, where the potential is roughly 
linear and qualitatively in agreement with Casimir scaling. 

\begin{table}[htb]
\vspace{-15pt}
\setlength{\tabcolsep}{.5pc}
\caption{Best estimate of string tensions in energy units from different
coupling constants.  The ratio of string 
tensions is proportional to the ratio of Casimir scaling of the last 
column.}
\label{k0-ener}
\begin{center}
\begin{tabular}{lccc}
\hline
Rep.  & K (GeV) & $\frac{K_{r}}{K_{f}}$ & $\frac{C_{r}}{C_{f}}$\\
\hline
3   &    $0.222(1)(8)(21)$  &    -           &    -           \\ \\
8   &    $0.437(2)(20)(42)$ & $1.97(1)(12)$  & $2.25$  \\ \\
6   &    $0.514(5)(67)(49)$ & $2.32(3)(31)$  & $2.5$    \\ \\
15a &    $0.77(1)(15)(7)$   & $3.47(5)(69)$  & $4.0$      \\ \\
10  &    $0.97(2)(27)(9)$   & $4.37(9)(123)$ & $4.5$      \\ \\
27  &    $1.12(1)(20)(11)$  & $5.05(5)(92)$  & $6$       \\ \\
15s &    $1.60(2)(52)(15)$  & $7.2(1)(24)$   & $7$       \\ 
\hline
\end{tabular}
\end{center}
\vspace{-20pt}
\end{table}

\begin{figure}[htb]
\vspace{35pt}
\epsfxsize=1. \hsize
\epsffile{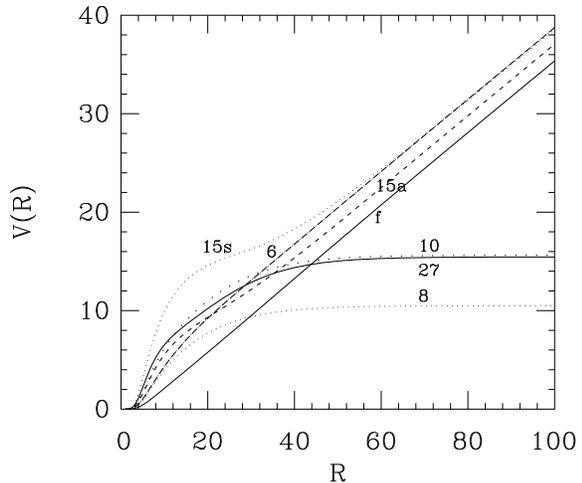}
\vspace{-57pt}
\caption{Potential between static sources for the range of 
$R \in [1,100]$. For
$R>40$, potentials for representations with zero triality are screened
and for the ones with non-zero triality closely parallel the fundamental
potential. The fundamental representation is shown by the letter ``f''. 
The scale of $R$ and $V(R)$ are arbitrary.}
\label{second}
\vspace{-15pt}
\end{figure}

The numerical lattice calculations of this work shows the existence of
a linear potential between static sources for fundamental and higher 
representations at small and intermediate distances. String tensions
roughly obey Casimir scaling. No screening or change of the slope 
is observed for higher representations. This is in contrast to the 
prediction of the fat-center-vortices model. Probably Wilson loops do not 
couple well to screened states and couple primarily to string-like, 
confined states. 

I thank my advisor Claude Bernard for his great support in 
this work. I wish to thank the MILC collaboration and specially Robert 
Sugar and Steven Gottlieb for computing resources.

\end{document}